\documentclass[prl,aps,twocolumn]{revtex4}
\usepackage{graphicx} 
\usepackage[usenames]{color}
\usepackage{amsmath,amssymb}
\usepackage{gensymb}
\usepackage{natbib}
\usepackage{xcolor}
\def\strutdepth{\dp\strutbox}
\def\nw#1{\strut\vadjust{\kern-\strutdepth\vtop to0pt{\vss\hbox to\hsize
{\hskip\hsize\hskip5pt$\leftarrow$\hss\strut}}}{\em #1}}

\DeclareMathOperator{\cotan}{cotan}

\begin{document}

\title{Contact angles on a soft solid: from Young's law to Neumann's law}
\author{Antonin Marchand$^1$, Siddhartha Das$^2$, Jacco H. Snoeijer$^3$ and Bruno Andreotti$^1$.}
\affiliation{
$^{1}$Physique et M\'ecanique des Milieux H\'et\'erog\`enes, UMR
7636 ESPCI -CNRS, Univ. Paris-Diderot, 10 rue Vauquelin, 75005, Paris\\
$^{2}$Department of Mechanical Engineering, University of Alberta, Canada T6G~2G8\\
$^{3}$Physics of Fluids Group and MESA+ Institute for Nanotechnology, 
University of Twente, P.O. Box 217, 7500 AE Enschede, The Netherlands
}

\date{\today}%

\begin{abstract}
The contact angle that a liquid drop makes on a soft substrate does not obey the classical Young's relation, since the solid is deformed elastically by the action of the capillary forces. The finite elasticity of the solid also renders the contact angles different from that predicted by Neumann's law, which applies when the drop is floating on another liquid. Here we derive an elasto-capillary model for contact angles on a soft solid, by coupling a mean-field model for the molecular interactions to elasticity. We demonstrate that the limit of vanishing elastic modulus yields Neumann's law or a slight variation thereof, depending on the force transmission in the solid surface layer. The change in contact angle from the rigid limit (Young) to the soft limit (Neumann) appears when the length scale defined by the ratio of surface tension to elastic modulus $\gamma/E$ reaches a few molecular sizes.
\end{abstract}

\maketitle

The wetting of liquid drops on deformable solids is important in many circumstances, with examples from biology to microfluidic devices~\cite{BicoNAT2004,HonsAPL2010,JungPOF2009,DupNAT2012}. When the solid is soft or flexible, the shape of both the solid and the liquid are determined by elasto-capillary interactions, i.e. by the elastic response to the capillary forces~\cite{PyPRL,BicoJPCM2010}. Till date, however, the most basic characterization of wetting has remained elusive for highly deformable solids \cite{RusanovUSSR,YukJCIS1986,ShanJPD1987,WhiteJCIS2003}: What is the contact angle that a liquid makes on a soft solid? 

The geometry of the interfaces near the three-phase contact line is governed by two classical laws that describe the macroscopic boundary condition for the contact angles~\cite{deGennes}. Young's law applies in the case where the substrate is perfectly rigid, with elastic modulus $E=\infty$, while Neumann's law holds for liquid lenses floating on another liquid ``substrate''. A question that naturally arises is whether the contact angles vary from ``Young" to ``Neumann" upon reducing the elastic modulus of the substrate: in other words, does one recover Neumann's angles in the limit $E\rightarrow 0$? Interestingly, the ratio of liquid-vapor surface tension $\gamma$ to elastic modulus $E$ has the dimension of a length. It has remained an object of discussion whether, for the solid to become highly deformable, this elastic length $\gamma/E$ should be comparable to a molecular size ~\cite{LesJCIS1961,DasPOF2011} or to a macroscopic length such as the size of the drop~\cite{JeriPRL2011,Sty2012}.

The difficulty of the problem results from its inherently multi-scale nature. On one hand, the capillary forces are localized in the vicinity of the contact-line. On the other hand, the Green function giving the surface displacement $\delta h(x)$, induced by a Dirac force distribution of resultant $f_z$ applied at the boundary of a two-dimensional elastic medium, scales as~\cite{Johnson}
\begin{equation}\label{eq:2D}
\delta h(x) \sim -\frac{f_z}{E} \, \ln |x|,
\end{equation}
and is therefore singular at both small and large distance $x$ from the contact line. An outer cut-off is naturally provided by the thickness $h$ of the elastic film or the size of the drop~\cite{LesJCIS1961,SroPRE1994,YuJCIS2009}. The inner regularization is commonly assumed to originate from the finite range of intermolecular capillary forces~\cite{RusanovUSSR,WhiteJCIS2003,DasPOF2011}, or by the breakdown of linear elasticity~\cite{TurBJ1999}. Hence, the transition from Young's to Neumann's contact angles calls for an unprecedented, fully self-consistent treatment of elastic and capillary interactions.

\begin{figure*}[t!]
\centering
\includegraphics{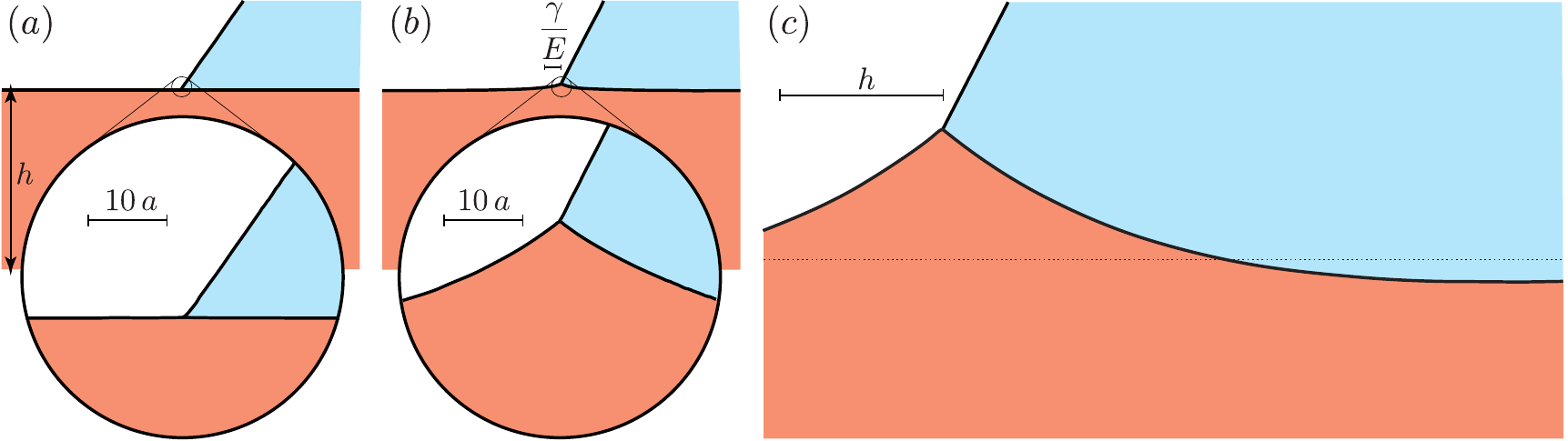}
\vspace{-3 mm}
\caption{Geometry near the three-phase contact line obtained by coupling elasticity to a DFT model. Contact angles continuously vary from Young's law to Neumann's law by reducing the stiffness of the solid. (a) Rigid solid, $\gamma/(Ea) \ll 1$. The surface is undeformed and the liquid contact angle follows Young's law down to molecular scale $a$. (b) Soft solid, $ \gamma/(Ea)\gg 1$. Surface elasticity is negligible on the scale of molecular interactions, and the contact angles obey Neumann's law.  The solid is deformed over a distance $\sim \gamma/E$ from the contact line. (c) Very soft solid, $\gamma/(Eh) \sim 1$. The change of the contact angles saturates when $\gamma/E$ becomes comparable to the thickness of the elastic film. The solid angle measured at scale $h$ becomes identical to the microscopic solid angle at scale $a$.}
\vspace{-3 mm}
\label{fig.sketch}
\end{figure*}

In this Letter, we solve the elasto-capillary contact angle selection within the framework of the Density Functional Theory, using the sharp-kink approximation. The evolution of the contact angles with stiffness is summarized in Fig.~\ref{fig.sketch}. The central result is that the liquid contact angle is selected at the molecular scale $a$ and therefore exhibits a transition from ``Young'' to ``Neumann''  around a dimensionless number $\gamma/(Ea)$ of order unity. We propose an analytical description of this transition, which agrees quantitatively with the full numerical solution of the coupled DFT and elasticity models. Above this transition, the elastic solid is deformed by the capillary forces over the length $\gamma/E$. When the latter becomes larger than the system size (the layer thickness $h$ in Fig.~\ref{fig.sketch}c), the elastic deformation saturates. 

\textit{Density Functional Theory~--~}The multi-scale nature of elasticity makes it convenient to treat the wetting interactions in a continuum framework, such as the DFT. We consider a simplified DFT model in which the solid and the liquid are treated as homogeneous phases that mutually attract, while the interface is assumed to be infinitely thin~\cite{GettaPRE1998,MerchantPFA1992,SnoeijerPOF2008}. This model captures the microscopic properties such as the stress-anisotropy near the interface, the disjoining pressure and the line tension and is consistent with macroscopic thermodynamics in the form of Laplace pressure and Young's law~\cite{GettaPRE1998,MerchantPFA1992,SnoeijerPOF2008,WeijsPOF2011}. 

The idea underlying this DFT model is to separate the molecular interactions into a long-range attractive potential $\varphi(r)$, which takes into account the pair correlation function, and a hard core repulsion that acts as a contact force. For van der Waals interactions this potential decays as $1/r^6$, which is cut-off at a microscopic distance $r = a$ that corresponds to the repulsive core. In the model, it turns out that all the capillary forces can be expressed in terms of the integrated potential~\cite{DasPOF2011},
\begin{equation}\label{eq:potential}
\Phi_{\alpha \beta} \left(\mathbf{r} \right) = \rho_\alpha \rho_\beta 
\int_\alpha \mathbf{dr'}\, \varphi_{\alpha\beta} \left(|\mathbf{r} - \mathbf{r'} |\right).
\end{equation}

This represents the potential energy in phase $\beta$ due to phase $\alpha$, where the phases can be liquid ($L$), solid ($S$) or vapor ($V$). $\rho_\alpha$ and $\rho_\beta$ are the corresponding homogeneous densities. The repulsive core at $r=a$ ensures that the integrals over the entire domain $\alpha$ converge, and is modeled by an isotropic internal pressure that ensures incompressibility. As detailed in \cite{DasPOF2011,sup}, the model distinguishes three types of attractive interactions: liquid-liquid, solid-solid and solid-liquid interactions, which can be expressed directly in terms of the surface tensions $\gamma$, $\gamma_{SV}$ and $\gamma_{SL}$~\cite{DasPOF2011,MerchantPFA1992,BauerEPJB1999}. The liquid-vapor surface tension $\gamma$ characterises the liquid-liquid interactions. The strength of the solid-liquid interactions is characterized by Young's contact angle $\theta_Y$, defined by $\cos \theta_Y=(\gamma_{SV}- \gamma_{SL})/\gamma$. The interaction with vapor can be neglected in the limit of a low vapor density. In the full DFT numerical calculation, the equilibrium shape of the liquid-vapor interface is determined iteratively using the procedure described in previous papers \cite{SnoeijerPOF2008,WeijsPOF2011}.

\begin{figure}[t!]
\includegraphics{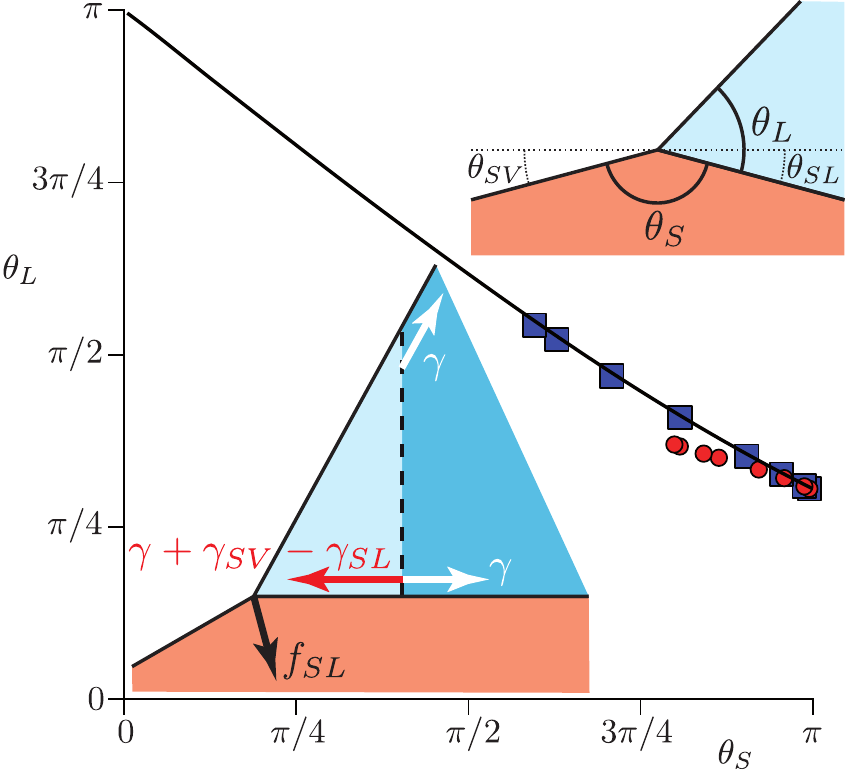}
\vspace{-3 mm}
\caption{(Color online) Main graph: Relation between $\theta_L$ and $\theta_S$ predicted by the DFT model. The solid line is the analytic formula (\ref{eq:thetaL}) for $\theta_Y = 0.96$. Symbols are the angles obtained numerically for the normal force transmission model ($\blacksquare$) and the vectorial force transmission model ($\bullet$), as defined in the text. Upper inset: definition of $\theta_L$ and $\theta_S$. Lower inset: forces acting on a corner of liquid (bright region, light blue). Black: force exerted by the solid. Red: repulsive liquid-liquid force induced by the presence of the solid. White: attractive force exerted by the liquid, due to the missing half domain of liquid.}
\vspace{-3 mm}
\label{fig.LiquidWedge}
\end{figure}

\textit{Selection of the liquid angle~--~}An important feature is that the strength of the capillary interactions depends on the geometry of the deformable solid. We consider the reference case of a solid shaped like a wedge of angle $\theta_S$ (upper inset of Fig.~\ref{fig.LiquidWedge}). Similar to the case of a flat surface, the force acting on a corner of liquid depends only on its angle $\theta_L$ at a large distance from the contact line, and can be determined exactly by integrating over all the interactions in the DFT model~\cite{MerchantPFA1992,SnoeijerPOF2008}. This force on the liquid corner consists of three contributions that are sketched in the lower inset of Fig.~\ref{fig.LiquidWedge}: (i) the force exerted by the solid (solid-liquid interactions, black arrow), (ii) the attractive force exerted by the rest of the liquid (liquid-liquid interactions, white arrows), and (iii) the repulsive force exerted by the rest of the liquid, induced by the presence of the solid~\cite{Nijm1990} (liquid-liquid interactions, red arrow). This last force arises because the presence of the solid leads to an increase of the liquid internal pressure near the solid-liquid interface.

The balance of forces in Fig.~\ref{fig.LiquidWedge} provides the equilibrium $\theta_L$ for arbitrary $\theta_S$ (details are worked out in the Supplementary~\cite{sup}):
\begin{eqnarray}\label{eq:thetaL}
\cos\theta_L = \frac12\left[ \cos\theta_Y[1-\cos\theta_S] \right.\nonumber \\
\left. -\sin\theta_S \sqrt{ \frac2{1-\cos\theta_S} -\cos^2\theta_Y } \right].
\end{eqnarray}
This result is independent of the microscopic length $a$ and the functional form of $\varphi(r)$. For a flat surface ($\theta_S=\pi$), the solid-on-liquid force is oriented vertically, with $f_{SL}=\gamma \sin \theta_L$. In this case the force balance reduces to Young's law, and the liquid angle $\theta_L=\theta_Y$. However, (\ref{eq:thetaL}) predicts that $\theta_L$ increases when $\theta_S$ is reduced (Fig.~\ref{fig.LiquidWedge}, solid line). Physically, this is due to the reduction of the solid volume for smaller $\theta_S$: this lowers the total solid-liquid interaction, making the solid wedge more ``hydrophobic''.

\textit{Selection of the solid angle~--~}If the phase $S$ behaves as a perfect liquid, its mechanical equilibrium gives a second equation for the angles. This can be deduced from (\ref{eq:thetaL}) by exchanging the roles of $L$ and $S$, which indeed result into $\theta_S$ and $\theta_L$ according to Neumann's law~\cite{sup}. In the elasto-capillary problem, by contrast, the solid $S$ can resist shear. One therefore needs to express how the capillary stress $\boldsymbol\sigma$ applied at the free surface  deforms the solid. We treat the substrate as an incompressible elastic body (Poisson ratio $\nu=1/2$) with Young's modulus $E$, as is typical for soft elastomers. Introducing the Green's function $\mathbf{R}$, which depends on the elastic properties and the geometry of the substrate, we get the surface displacement:
\begin{equation}\nonumber
\delta h(x)=\frac{1}{E} \int_{-\infty}^\infty \mathbf{R}(x-x';h) \cdot \boldsymbol\sigma(x')dx'.
\end{equation}
The contact line is considered to be invariant in one direction, so that  $\mathbf{R}$ and $\boldsymbol \sigma$ have two components corresponding to the normal and the tangential directions to the substrate. The elastic kernel requires a cut-off length at large scale, which for our numerical calculations arises due to the finite elastic film thickness $h$~\cite{JeriPRL2011}. The capillarity-induced $\boldsymbol \sigma$ can be expressed in terms of the $\Phi_{\alpha \beta}$~\cite{sup}, and the integrals of (\ref{eq:potential}) can be evaluated numerically for arbitrary shape of the liquid and solid domains. This closes the elasto-capillary problem and the resulting numerical profiles are provided in Fig.~\ref{fig.sketch}. 

At intermediate distances from the contact line, $a \ll x \ll h$, the Green's function for the elastic response is given by Eq.~(\ref{eq:2D}). The slope of the solid-liquid interface thus scales as $\delta h' \sim f_z /(Ex)$.  Importantly, the angle $\theta_L$ of the liquid is selected at the micro-scale $a$. Therefore, the relevant solid angle $\theta_S$ induced by elastic deformations must be defined at that scale. This is confirmed by the agreement between the prediction of (\ref{eq:thetaL}) and the numerical solution of the fully coupled elasticity-DFT model: the symbols in Fig.~\ref{fig.LiquidWedge} are obtained by measuring $\theta_S$ in the numerics at a distance $a$ from the contact line. With this information, one can obtain an approximate equation for the selection of $\theta_S$ by evaluating (\ref{eq:2D}) at $x=a$:
\begin{equation}\label{eq:alphabeta}
\delta h' \sim \tan\left(\frac{\pi-\theta_S}{2}\right)  \sim \frac{f_z}{Ea}.
\end{equation}

\begin{figure}[t!]
\includegraphics{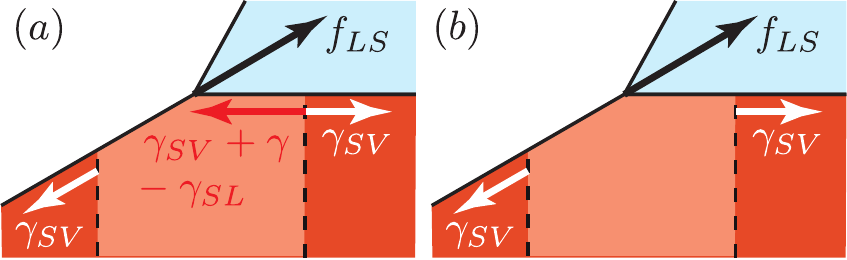}
\vspace{-3 mm}
\caption{(Color online) Forces acting on the corner of solid near the contact line (indicated by the bright (light orange) region near the contact line). (a) Normal force transmission model. Black: force exerted by the liquid. Red: force exerted by the solid due to pressure-build induced by the liquid. White: force exerted by the solid, due to the missing half domain of solid. (b) Vectorial force transmission model. The difference with respect to (a) is the absence of pressure build-up in the surface layer (red).}
\vspace{-3 mm}
\label{fig.SolidWedgeSchematic}
\end{figure}

\textit{The force acting on the solid corner.~--~}The final step is to express the vertical force $f_z$ exerted on the solid corner in the vicinity of the contact line (bright, light orange region in Fig.~\ref{fig.SolidWedgeSchematic}). Using the approximation that the solid domain is a perfect wedge and assuming that the liquid is at equilibrium, we can derive the tangential and normal components of this force due to the liquid-solid interaction~\cite{sup},
\begin{eqnarray}
\frac{f_{LS}^t}{\gamma}&=&(1+\cos\theta_Y) \frac{\cos\frac{\theta_L}2\,\sin\frac{\theta_S}2}{\sin\frac{\theta_L+\theta_S}2}\;,\\
\frac{f_{LS}^n}{\gamma}&=&\frac{(1+\cos\theta_{Y})}2\left(\sin\theta_S+\frac{\cos \theta_S}{\tan\frac{\theta_L+\theta_S}{2}}\right) 
\nonumber \\ 
&&
+\frac{(1-\cos\theta_{Y})}2\cotan\frac{\theta_{L}}2\;.
\label{eq:ftfn}
\end{eqnarray}
As emphasized in recent papers, this force is oriented towards the interior of the liquid and therefore presents a large tangential component, even in the limit of a flat surface~\cite{DasPOF2011,MarPRL2012}.

To express the solid-solid interactions, we need to model the mechanical behavior of the surface layer of the substrate. We consider two extreme cases of how the liquid-on-solid force can be transmitted to the bulk of the elastic solid. First, one can assume that only the normal stress is transmitted, as would be the case for a liquid. In terms of forces on the solid corner (bright region in Fig.~\ref{fig.SolidWedgeSchematic}), the tangential component of $f_{LS}$ is balanced by a pressure build-up in the surface layer. This is represented by the red arrow in Fig.~\ref{fig.SolidWedgeSchematic}a (in perfect analogy to the red arrow in the liquid in Fig.~\ref{fig.LiquidWedge}). In this case of \emph{normal force transmission}, the total vertical force reads
\begin{equation}
f_z=f_{LS}^n\cos\theta_{SL}+(-f_{LS}^t+\gamma-\gamma_{SL})\,\sin \theta_{SL}-\gamma_{\rm SV}\,\sin \theta_{SV}\;.\label{eq:fzsl}
\end{equation}
The angles $\theta_{SL}$ and $\theta_{SV}$ are defined with respect to the undisturbed solid surface (Fig.~\ref{fig.LiquidWedge}).

Alternatively, one can hypothesize a perfect \emph{vectorial force transmission}, for which there is no such pressure build-up in the surface layer (Fig.~\ref{fig.SolidWedgeSchematic}b). We recently proposed an experimental test aiming to discriminate between the two force transmission models: it turned out that the vectorial transmission model is the correct description for an elastomer~\cite{MarPRL2012}. Then, the tangential force exerted by the liquid is transmitted to the bulk of the elastic body, and the total force on the solid corner becomes (Fig.~\ref{fig.SolidWedgeSchematic}b)
\begin{equation}
f_z=f_{LS}^n\cos\theta_{SL}+(-f_{LS}^t-\gamma_{SV})\,\sin \theta_{SL}-\gamma_{\rm SV}\,\sin \theta_{SV}\;.\label{eq:fzsl2}
\end{equation}
\begin{figure}[t!]
\includegraphics{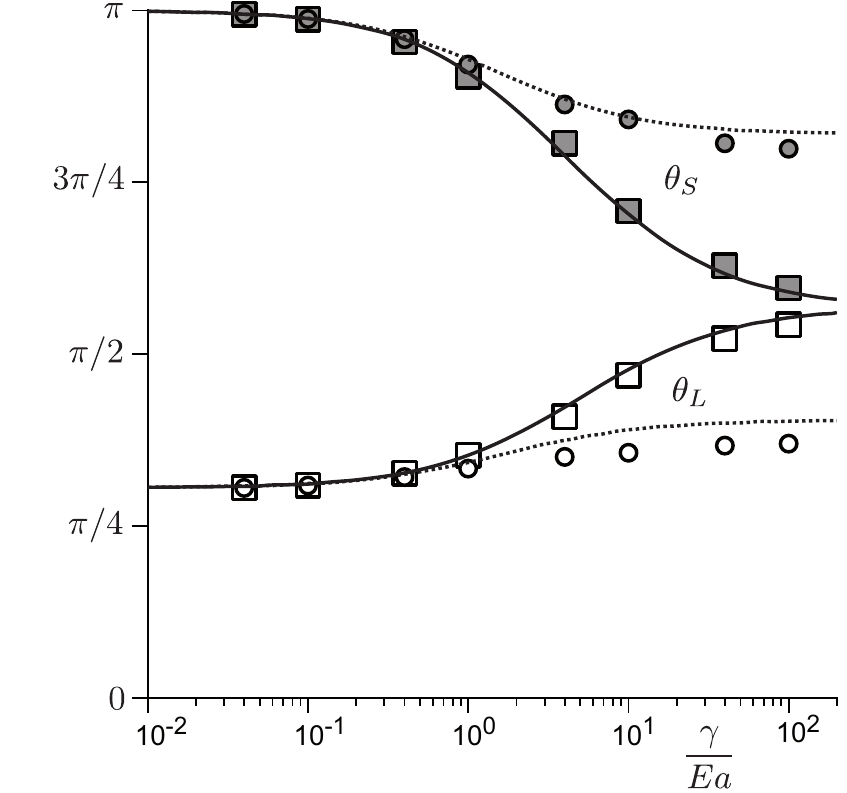}
\vspace{-3 mm}
\caption{Transition of the contact angles $\theta_L$ (white), $\theta_S$ (gray) upon increasing the ``softness'' parameter  $\gamma/(Ea)$ for $\theta_Y=0.96$ and $\gamma_{SV}=\gamma$. Symbols correspond to DFT numerical solutions  for $h/a=1000$ for the normal transmission force ($\square$) and for the vectorial transmission force ($\circ$).}
\vspace{-3 mm}
\label{fig.angles}
\end{figure}

\textit{Discussion~--~}The system of equations (\ref{eq:thetaL}--\ref{eq:ftfn}), closed by Eq.~(\ref{eq:fzsl}) or Eq.~(\ref{eq:fzsl2}), gives a prediction for $\theta_L$ and $\theta_S$, and forms the central result of our paper. It involves three dimensionless parameters: the elasto-capillary number $\gamma/(Ea)$ and the surface tension ratios $\gamma_{SV}/\gamma$, $\gamma_{SL}/\gamma$. The resulting contact angles are shown as solid lines in Fig.~\ref{fig.angles}. The symbols provide a direct comparison with numerical solution of the full elasticity-DFT model.

In the limit of strong elasticity, $\gamma/(Ea) \ll 1$, one recovers Young's angle $\theta_L=\theta_Y$ and an undeformed solid $\theta_S=\pi$, regardless of the model of force transmission. In the opposite limit of a soft solid, $\gamma/(Ea) \gg 1$, the elasticity is too weak to resist any force near the contact line. In this limit one thus finds that $f_z =0$, corresponding to a perfect balance of capillary forces. Elementary algebra shows that for the \emph{normal force transmission} model (Eq.~\ref{eq:fzsl}), the angles $\theta_L,\theta_S$ perfectly correspond to Neumann's triangle in the soft limit.  Treating the capillary force as perfectly localized~\cite{Sty2012}, i.e. $a=0$, corresponds to this limiting case. However, a continuous transition between ``Young'' and ``Neumann'' is observed in Fig.~\ref{fig.angles}, with a crossover around $\gamma/(Ea) \sim 1$. A similar transition of the contact angles is observed for the \emph{vectorial force transmission} model (Eq.~\ref{eq:fzsl2}), but with angles in the soft limit different from Neumann's law. 

Capillarity induced surface deformations are known to have major consequences for condensation of drops on soft solids \cite{SokulerLANGMUIR}, as used for micropatterning of polymeric surfaces \cite{Shojaei2009}, mechanical stability of gels \cite{MoraPRL2010}, or wetting on very soft coatings. The presented generalization of the classical laws for the contact angle will provide the essential boundary condition for such problems. 



\begin{thebibliography}{30}
\expandafter\ifx\csname natexlab\endcsname\relax\def\natexlab#1{#1}\fi
\expandafter\ifx\csname bibnamefont\endcsname\relax
  \def\bibnamefont#1{#1}\fi
\expandafter\ifx\csname bibfnamefont\endcsname\relax
  \def\bibfnamefont#1{#1}\fi
\expandafter\ifx\csname citenamefont\endcsname\relax
  \def\citenamefont#1{#1}\fi
\expandafter\ifx\csname url\endcsname\relax
  \def\url#1{\texttt{#1}}\fi
\expandafter\ifx\csname urlprefix\endcsname\relax\def\urlprefix{URL }\fi
\providecommand{\bibinfo}[2]{#2}
\providecommand{\eprint}[2][]{\url{#2}}

\bibitem[{\citenamefont{Bico et~al.}({2004})\citenamefont{Bico, Roman, Moulin,
  and Boudaoud}}]{BicoNAT2004}
\bibinfo{author}{\bibfnamefont{J.}~\bibnamefont{Bico}},
  \bibinfo{author}{\bibfnamefont{B.}~\bibnamefont{Roman}},
  \bibinfo{author}{\bibfnamefont{L.}~\bibnamefont{Moulin}}, \bibnamefont{and}
  \bibinfo{author}{\bibfnamefont{A.}~\bibnamefont{Boudaoud}},
  \bibinfo{journal}{{Nature}} \textbf{\bibinfo{volume}{{432}}},
  \bibinfo{pages}{{690}} (\bibinfo{year}{{2004}}).

\bibitem[{\citenamefont{van Honschoten et~al.}({2010})\citenamefont{van
  Honschoten, Berenschot, Ondaruhu, Sanders, Sundaram, Elwenspoek, and
  Tas}}]{HonsAPL2010}
\bibinfo{author}{\bibfnamefont{J.~W.} \bibnamefont{van Honschoten}},
  \bibinfo{author}{\bibfnamefont{J.~W.} \bibnamefont{Berenschot}},
  \bibinfo{author}{\bibfnamefont{T.}~\bibnamefont{Ondaruhu}},
  \bibinfo{author}{\bibfnamefont{R.~G.~P.} \bibnamefont{Sanders}},
  \bibinfo{author}{\bibfnamefont{J.}~\bibnamefont{Sundaram}},
  \bibinfo{author}{\bibfnamefont{M.}~\bibnamefont{Elwenspoek}},
  \bibnamefont{and} \bibinfo{author}{\bibfnamefont{N.~R.} \bibnamefont{Tas}},
  \bibinfo{journal}{App. Phys. Lett.} \textbf{\bibinfo{volume}{{97}}},
  \bibinfo{pages}{{014103}} (\bibinfo{year}{{2010}}).

\bibitem[{\citenamefont{Jung et~al.}({2009})\citenamefont{Jung, Reis, James,
  Clanet, and Bush}}]{JungPOF2009}
\bibinfo{author}{\bibfnamefont{S.}~\bibnamefont{Jung}},
  \bibinfo{author}{\bibfnamefont{P.~M.} \bibnamefont{Reis}},
  \bibinfo{author}{\bibfnamefont{J.}~\bibnamefont{James}},
  \bibinfo{author}{\bibfnamefont{C.}~\bibnamefont{Clanet}}, \bibnamefont{and}
  \bibinfo{author}{\bibfnamefont{J.~W.~M.} \bibnamefont{Bush}},
  \bibinfo{journal}{{Phys. Fluids}} \textbf{\bibinfo{volume}{{21}}},
  \bibinfo{pages}{{091110}} (\bibinfo{year}{{2009}}).

\bibitem[{\citenamefont{Duprat et~al.}({2012})\citenamefont{Duprat, Protiere,
  Beebe, and Stone}}]{DupNAT2012}
\bibinfo{author}{\bibfnamefont{C.}~\bibnamefont{Duprat}},
  \bibinfo{author}{\bibfnamefont{S.}~\bibnamefont{Protiere}},
  \bibinfo{author}{\bibfnamefont{A.~Y.} \bibnamefont{Beebe}}, \bibnamefont{and}
  \bibinfo{author}{\bibfnamefont{H.~A.} \bibnamefont{Stone}},
  \bibinfo{journal}{{Nature}} \textbf{\bibinfo{volume}{{482}}},
  \bibinfo{pages}{{510}} (\bibinfo{year}{{2012}}).

\bibitem[{\citenamefont{Py et~al.}({2007})\citenamefont{Py, Reverdy, Doppler,
  Bico, Roman, and Baroud}}]{PyPRL}
\bibinfo{author}{\bibfnamefont{C.}~\bibnamefont{Py}},
  \bibinfo{author}{\bibfnamefont{P.}~\bibnamefont{Reverdy}},
  \bibinfo{author}{\bibfnamefont{L.}~\bibnamefont{Doppler}},
  \bibinfo{author}{\bibfnamefont{J.}~\bibnamefont{Bico}},
  \bibinfo{author}{\bibfnamefont{B.}~\bibnamefont{Roman}}, \bibnamefont{and}
  \bibinfo{author}{\bibfnamefont{C.~N.} \bibnamefont{Baroud}},
  \bibinfo{journal}{Phys. Rev. Lett.} \textbf{\bibinfo{volume}{{98}}},
  \bibinfo{pages}{{156103}} (\bibinfo{year}{{2007}}).

\bibitem[{\citenamefont{Bico and Roman}({2010})}]{BicoJPCM2010}
\bibinfo{author}{\bibfnamefont{J.}~\bibnamefont{Bico}} \bibnamefont{and}
  \bibinfo{author}{\bibfnamefont{B.}~\bibnamefont{Roman}}, \bibinfo{journal}{J.
  Phys. Cond. Mat.} \textbf{\bibinfo{volume}{{22}}}, \bibinfo{pages}{{493101}}
  (\bibinfo{year}{{2010}}).

\bibitem[{\citenamefont{Rusanov}({1975})}]{RusanovUSSR}
\bibinfo{author}{\bibfnamefont{A.~I.} \bibnamefont{Rusanov}},
  \bibinfo{journal}{{Coll. J. USSR}} \textbf{\bibinfo{volume}{{37}}},
  \bibinfo{pages}{{614}} (\bibinfo{year}{{1975}}).

\bibitem[{\citenamefont{Yuk and Jhon}({1986})}]{YukJCIS1986}
\bibinfo{author}{\bibfnamefont{S.~H.} \bibnamefont{Yuk}} \bibnamefont{and}
  \bibinfo{author}{\bibfnamefont{M.~S.} \bibnamefont{Jhon}},
  \bibinfo{journal}{J. Coll. Int. Sci.} \textbf{\bibinfo{volume}{{110}}},
  \bibinfo{pages}{{252}} (\bibinfo{year}{{1986}}).

\bibitem[{\citenamefont{Shanahan}({1987})}]{ShanJPD1987}
\bibinfo{author}{\bibfnamefont{M.~E.~R.} \bibnamefont{Shanahan}},
  \bibinfo{journal}{J. Phys. D App. Phys.} \textbf{\bibinfo{volume}{{20}}},
  \bibinfo{pages}{{945}} (\bibinfo{year}{{1987}}).

\bibitem[{\citenamefont{White}({2003})}]{WhiteJCIS2003}
\bibinfo{author}{\bibfnamefont{L.~R.} \bibnamefont{White}},
  \bibinfo{journal}{{J. Coll. Int. Sci.}} \textbf{\bibinfo{volume}{{258}}},
  \bibinfo{pages}{{82}} (\bibinfo{year}{{2003}}).

\bibitem[{\citenamefont{de~Gennes et~al.}(2003)\citenamefont{de~Gennes,
  Brochard-Wyart, and Quere}}]{deGennes}
\bibinfo{author}{\bibfnamefont{P.-G.} \bibnamefont{de~Gennes}},
  \bibinfo{author}{\bibfnamefont{F.}~\bibnamefont{Brochard-Wyart}},
  \bibnamefont{and} \bibinfo{author}{\bibfnamefont{D.}~\bibnamefont{Quere}},
  \emph{\bibinfo{title}{Capillarity and Wetting Phenomena: Drops, Bubbles,
  Pearls, Waves}} (\bibinfo{publisher}{Springer, New York},
  \bibinfo{year}{2003}).

\bibitem[{\citenamefont{Lester}({1961})}]{LesJCIS1961}
\bibinfo{author}{\bibfnamefont{G.~R.} \bibnamefont{Lester}},
  \bibinfo{journal}{{J. Coll. Int. Sci.}} \textbf{\bibinfo{volume}{{16}}},
  \bibinfo{pages}{{315}} (\bibinfo{year}{{1961}}).

\bibitem[{\citenamefont{Das et~al.}({2011})\citenamefont{Das, Marchand,
  Andreotti, and Snoeijer}}]{DasPOF2011}
\bibinfo{author}{\bibfnamefont{S.}~\bibnamefont{Das}},
  \bibinfo{author}{\bibfnamefont{A.}~\bibnamefont{Marchand}},
  \bibinfo{author}{\bibfnamefont{B.}~\bibnamefont{Andreotti}},
  \bibnamefont{and} \bibinfo{author}{\bibfnamefont{J.~H.}
  \bibnamefont{Snoeijer}}, \bibinfo{journal}{{Phys. Fluids}}
  \textbf{\bibinfo{volume}{{23}}}, \bibinfo{pages}{{072006}}
  (\bibinfo{year}{{2011}}).

\bibitem[{\citenamefont{Jerison et~al.}({2011})\citenamefont{Jerison, Xu,
  Wilen, and Dufresne}}]{JeriPRL2011}
\bibinfo{author}{\bibfnamefont{E.~R.} \bibnamefont{Jerison}},
  \bibinfo{author}{\bibfnamefont{Y.}~\bibnamefont{Xu}},
  \bibinfo{author}{\bibfnamefont{L.~A.} \bibnamefont{Wilen}}, \bibnamefont{and}
  \bibinfo{author}{\bibfnamefont{E.~R.} \bibnamefont{Dufresne}},
  \bibinfo{journal}{{Phys. Rev. Lett.}} \textbf{\bibinfo{volume}{{106}}},
  \bibinfo{pages}{{186103}} (\bibinfo{year}{{2011}}).

\bibitem[{\citenamefont{Style and Dufresne}({2012})}]{Sty2012}
\bibinfo{author}{\bibfnamefont{R.~W.} \bibnamefont{Style}} \bibnamefont{and}
  \bibinfo{author}{\bibfnamefont{E.~R.} \bibnamefont{Dufresne}},
  \bibinfo{journal}{arXiv:} p. \bibinfo{pages}{{1203.1654}}
  (\bibinfo{year}{{2012}}).

\bibitem[{\citenamefont{Johnson}(1985)}]{Johnson}
\bibinfo{author}{\bibfnamefont{K.~L.} \bibnamefont{Johnson}},
  \emph{\bibinfo{title}{Contact Mechanics}} (\bibinfo{publisher}{Cambridge
  University Press, Cambridge}, \bibinfo{year}{1985}).

\bibitem[{\citenamefont{Srolovitz et~al.}({1994})\citenamefont{Srolovitz,
  Safran, and Tenne}}]{SroPRE1994}
\bibinfo{author}{\bibfnamefont{D.~J.} \bibnamefont{Srolovitz}},
  \bibinfo{author}{\bibfnamefont{S.~A.} \bibnamefont{Safran}},
  \bibnamefont{and} \bibinfo{author}{\bibfnamefont{R.}~\bibnamefont{Tenne}},
  \bibinfo{journal}{{Phys. Rev. E}} \textbf{\bibinfo{volume}{{49}}},
  \bibinfo{pages}{{5260}} (\bibinfo{year}{{1994}}).

\bibitem[{\citenamefont{Yu and Zhao}({2009})}]{YuJCIS2009}
\bibinfo{author}{\bibfnamefont{Y.-S.} \bibnamefont{Yu}} \bibnamefont{and}
  \bibinfo{author}{\bibfnamefont{Y.-P.} \bibnamefont{Zhao}},
  \bibinfo{journal}{J. Coll. Int. Sci.} \textbf{\bibinfo{volume}{{339}}},
  \bibinfo{pages}{{489}} (\bibinfo{year}{{2009}}).

\bibitem[{\citenamefont{Turner and Sens}({1999})}]{TurBJ1999}
\bibinfo{author}{\bibfnamefont{M.~S.} \bibnamefont{Turner}} \bibnamefont{and}
  \bibinfo{author}{\bibfnamefont{P.}~\bibnamefont{Sens}},
  \bibinfo{journal}{Biophys. J} \textbf{\bibinfo{volume}{{76}}},
  \bibinfo{pages}{{564}} (\bibinfo{year}{{1999}}).

\bibitem[{\citenamefont{Getta and Dietrich}({1998})}]{GettaPRE1998}
\bibinfo{author}{\bibfnamefont{T.}~\bibnamefont{Getta}} \bibnamefont{and}
  \bibinfo{author}{\bibfnamefont{S.}~\bibnamefont{Dietrich}},
  \bibinfo{journal}{{Phys. Rev. E}} \textbf{\bibinfo{volume}{{57}}},
  \bibinfo{pages}{{655}} (\bibinfo{year}{{1998}}).

\bibitem[{\citenamefont{Merchant and Keller}({1992})}]{MerchantPFA1992}
\bibinfo{author}{\bibfnamefont{G.~J.} \bibnamefont{Merchant}} \bibnamefont{and}
  \bibinfo{author}{\bibfnamefont{J.~B.} \bibnamefont{Keller}},
  \bibinfo{journal}{Phys. Fluids A} \textbf{\bibinfo{volume}{{4}}},
  \bibinfo{pages}{{477}} (\bibinfo{year}{{1992}}).

\bibitem[{\citenamefont{Snoeijer and Andreotti}({2008})}]{SnoeijerPOF2008}
\bibinfo{author}{\bibfnamefont{J.~H.} \bibnamefont{Snoeijer}} \bibnamefont{and}
  \bibinfo{author}{\bibfnamefont{B.}~\bibnamefont{Andreotti}},
  \bibinfo{journal}{{Phys. Fluids}} \textbf{\bibinfo{volume}{{20}}},
  \bibinfo{pages}{{057101}} (\bibinfo{year}{{2008}}).

\bibitem[{\citenamefont{Weijs et~al.}({2011})\citenamefont{Weijs, Marchand,
  Andreotti, Lohse, and Snoeijer}}]{WeijsPOF2011}
\bibinfo{author}{\bibfnamefont{J.}~\bibnamefont{Weijs}},
  \bibinfo{author}{\bibfnamefont{A.}~\bibnamefont{Marchand}},
  \bibinfo{author}{\bibfnamefont{B.}~\bibnamefont{Andreotti}},
  \bibinfo{author}{\bibfnamefont{D.}~\bibnamefont{Lohse}}, \bibnamefont{and}
  \bibinfo{author}{\bibfnamefont{J.~H.} \bibnamefont{Snoeijer}},
  \bibinfo{journal}{Phys. Fluids} \textbf{\bibinfo{volume}{{23}}},
  \bibinfo{pages}{{022001}} (\bibinfo{year}{{2011}}).

\bibitem[{\citenamefont{Das et~al.}({2012})\citenamefont{Das, Marchand,
  Andreotti, and Snoeijer}}]{sup}
\bibinfo{author}{\bibfnamefont{S.}~\bibnamefont{Das}},
  \bibinfo{author}{\bibfnamefont{A.}~\bibnamefont{Marchand}},
  \bibinfo{author}{\bibfnamefont{B.}~\bibnamefont{Andreotti}},
  \bibnamefont{and} \bibinfo{author}{\bibfnamefont{J.~H.}
  \bibnamefont{Snoeijer}}, \bibinfo{journal}{{See Supplementary Material}}
  (\bibinfo{year}{{2012}}).

\bibitem[{\citenamefont{Bauer and Dietrich}({1999})}]{BauerEPJB1999}
\bibinfo{author}{\bibfnamefont{C.}~\bibnamefont{Bauer}} \bibnamefont{and}
  \bibinfo{author}{\bibfnamefont{S.}~\bibnamefont{Dietrich}},
  \bibinfo{journal}{{Eur. Phys. J. B}} \textbf{\bibinfo{volume}{{10}}},
  \bibinfo{pages}{{767}} (\bibinfo{year}{{1999}}).

\bibitem[{\citenamefont{Nijmeijer et~al.}({1990})\citenamefont{Nijmeijer,
  Bruin, Bakker, and Van~Leeuwen}}]{Nijm1990}
\bibinfo{author}{\bibfnamefont{J.}~\bibnamefont{Nijmeijer}},
  \bibinfo{author}{\bibfnamefont{C.}~\bibnamefont{Bruin}},
  \bibinfo{author}{\bibfnamefont{A.}~\bibnamefont{Bakker}}, \bibnamefont{and}
  \bibinfo{author}{\bibfnamefont{J.}~\bibnamefont{Van~Leeuwen}},
  \bibinfo{journal}{{Phys. Rev. A}} \textbf{\bibinfo{volume}{{42}}},
  \bibinfo{pages}{{6052}} (\bibinfo{year}{{1990}}).

\bibitem[{\citenamefont{Marchand et~al.}({2012})\citenamefont{Marchand, Das,
  Snoeijer, and Andreotti}}]{MarPRL2012}
\bibinfo{author}{\bibfnamefont{A.}~\bibnamefont{Marchand}},
  \bibinfo{author}{\bibfnamefont{S.}~\bibnamefont{Das}},
  \bibinfo{author}{\bibfnamefont{J.~H.} \bibnamefont{Snoeijer}},
  \bibnamefont{and}
  \bibinfo{author}{\bibfnamefont{B.}~\bibnamefont{Andreotti}},
  \bibinfo{journal}{Phys. Rev. Lett.} \textbf{\bibinfo{volume}{{108}}},
  \bibinfo{pages}{{094301}} (\bibinfo{year}{{2012}}).

\bibitem[{\citenamefont{Sokuler et~al.}({2010})\citenamefont{Sokuler,
  Auernhammer, Roth, Liu, Bonaccurso, and Butt}}]{SokulerLANGMUIR}
\bibinfo{author}{\bibfnamefont{M.}~\bibnamefont{Sokuler}},
  \bibinfo{author}{\bibfnamefont{G.~K.} \bibnamefont{Auernhammer}},
  \bibinfo{author}{\bibfnamefont{M.}~\bibnamefont{Roth}},
  \bibinfo{author}{\bibfnamefont{C.}~\bibnamefont{Liu}},
  \bibinfo{author}{\bibfnamefont{E.}~\bibnamefont{Bonaccurso}},
  \bibnamefont{and} \bibinfo{author}{\bibfnamefont{H.-J.} \bibnamefont{Butt}},
  \bibinfo{journal}{Langmuir} \textbf{\bibinfo{volume}{{26}}},
  \bibinfo{pages}{{1544}} (\bibinfo{year}{{2010}}).

\bibitem[{\citenamefont{Shojaei-Zadeh
  et~al.}({2009})\citenamefont{Shojaei-Zadeh, Swanson, and Anna}}]{Shojaei2009}
\bibinfo{author}{\bibfnamefont{S.}~\bibnamefont{Shojaei-Zadeh}},
  \bibinfo{author}{\bibfnamefont{S.}~\bibnamefont{Swanson}}, \bibnamefont{and}
  \bibinfo{author}{\bibfnamefont{S.~L.} \bibnamefont{Anna}},
  \bibinfo{journal}{{Soft Matter}} \textbf{\bibinfo{volume}{{5}}},
  \bibinfo{pages}{{743}} (\bibinfo{year}{{2009}}).

\bibitem[{\citenamefont{Mora et~al.}({2010})\citenamefont{Mora, Phou,
  Fromental, Pismen, and Pomeau}}]{MoraPRL2010}
\bibinfo{author}{\bibfnamefont{S.}~\bibnamefont{Mora}},
  \bibinfo{author}{\bibfnamefont{T.}~\bibnamefont{Phou}},
  \bibinfo{author}{\bibfnamefont{J.-M.} \bibnamefont{Fromental}},
  \bibinfo{author}{\bibfnamefont{L.~M.} \bibnamefont{Pismen}},
  \bibnamefont{and} \bibinfo{author}{\bibfnamefont{Y.}~\bibnamefont{Pomeau}},
  \bibinfo{journal}{{Phys. Rev. Lett.}} \textbf{\bibinfo{volume}{{105}}},
  \bibinfo{pages}{{214301}} (\bibinfo{year}{{2010}}).

\end{thebibliography}


\begin{thebibliography}{6}
\expandafter\ifx\csname natexlab\endcsname\relax\def\natexlab#1{#1}\fi
\expandafter\ifx\csname bibnamefont\endcsname\relax
  \def\bibnamefont#1{#1}\fi
\expandafter\ifx\csname bibfnamefont\endcsname\relax
  \def\bibfnamefont#1{#1}\fi
\expandafter\ifx\csname citenamefont\endcsname\relax
  \def\citenamefont#1{#1}\fi
\expandafter\ifx\csname url\endcsname\relax
  \def\url#1{\texttt{#1}}\fi
\expandafter\ifx\csname urlprefix\endcsname\relax\def\urlprefix{URL }\fi
\providecommand{\bibinfo}[2]{#2}
\providecommand{\eprint}[2][]{\url{#2}}

\bibitem[{\citenamefont{Snoeijer and Andreotti}({2008})}]{SnoeijerPOF2008}
\bibinfo{author}{\bibfnamefont{J.~H.} \bibnamefont{Snoeijer}} \bibnamefont{and}
  \bibinfo{author}{\bibfnamefont{B.}~\bibnamefont{Andreotti}},
  \bibinfo{journal}{{Phys. Fluid.}} \textbf{\bibinfo{volume}{\textbf{20}}},
  \bibinfo{pages}{{057101}} (\bibinfo{year}{{2008}}).

\bibitem[{\citenamefont{Das et~al.}({2011})\citenamefont{Das, Marchand,
  Andreotti, and Snoeijer}}]{DasPOF2011}
\bibinfo{author}{\bibfnamefont{S.}~\bibnamefont{Das}},
  \bibinfo{author}{\bibfnamefont{A.}~\bibnamefont{Marchand}},
  \bibinfo{author}{\bibfnamefont{B.}~\bibnamefont{Andreotti}},
  \bibnamefont{and} \bibinfo{author}{\bibfnamefont{J.~H.}
  \bibnamefont{Snoeijer}}, \bibinfo{journal}{{Phys. Fluid}}
  \textbf{\bibinfo{volume}{\textbf{23}}}, \bibinfo{pages}{{072006}}
  (\bibinfo{year}{{2011}}).

\bibitem[{\citenamefont{Israelachvili}(1992)}]{Israelachvili}
\bibinfo{author}{\bibfnamefont{J.}~\bibnamefont{Israelachvili}},
  \emph{\bibinfo{title}{{Intermolecular and surface forces}}}
  (\bibinfo{publisher}{Academic Press, London}, \bibinfo{year}{1992}).

\bibitem[{\citenamefont{Marchand et~al.}(2011)\citenamefont{Marchand, Weijs,
  Snoeijer, and Andreotti}}]{MarchandAJP11}
\bibinfo{author}{\bibfnamefont{A.}~\bibnamefont{Marchand}},
  \bibinfo{author}{\bibfnamefont{J.~H.} \bibnamefont{Weijs}},
  \bibinfo{author}{\bibfnamefont{J.~H.} \bibnamefont{Snoeijer}},
  \bibnamefont{and}
  \bibinfo{author}{\bibfnamefont{B.}~\bibnamefont{Andreotti}},
  \bibinfo{journal}{American Journal Of Physics} \textbf{\bibinfo{volume}{79}},
  \bibinfo{pages}{999} (\bibinfo{year}{2011}).

\bibitem[{\citenamefont{Neumann}(1894)}]{Neumann}
\bibinfo{author}{\bibfnamefont{F.}~\bibnamefont{Neumann}},
  \emph{\bibinfo{title}{{Vorlesungen {\"u}ber die Theorie der
  Capillarit{\"a}t}}} (\bibinfo{publisher}{Teubner, Leipzig},
  \bibinfo{year}{1894}).

\bibitem[{\citenamefont{Marchand et~al.}({2012})\citenamefont{Marchand, Das,
  Snoeijer, and Andreotti}}]{MarPRL2012}
\bibinfo{author}{\bibfnamefont{A.}~\bibnamefont{Marchand}},
  \bibinfo{author}{\bibfnamefont{S.}~\bibnamefont{Das}},
  \bibinfo{author}{\bibfnamefont{J.~H.} \bibnamefont{Snoeijer}},
  \bibnamefont{and}
  \bibinfo{author}{\bibfnamefont{B.}~\bibnamefont{Andreotti}},
  \bibinfo{journal}{Phys. Rev. Lett.} \textbf{\bibinfo{volume}{\textbf{108}}},
  \bibinfo{pages}{{094301}} (\bibinfo{year}{{2012}}).

\end{thebibliography}
\end{document}


\title{Contact angles on a soft solid: from Young's law to Neumann's law\\(Supplementary Material)}
\author{Antonin Marchand$^1$, Siddhartha Das$^2$, Jacco H. Snoeijer$^3$ and Bruno Andreotti$^1$.}
\affiliation{
$^{1}$Physique et M\'ecanique des Milieux H\'et\'erog\`enes, UMR
7636 ESPCI -CNRS, Univ. Paris-Diderot, 10 rue Vauquelin, 75005, Paris\\
$^{2}$Department of Mechanical Engineering, University of Alberta, Canada T6G~2G8\\
$^{3}$Physics of Fluids Group and MESA+ Institute for Nanotechnology, 
University of Twente, P.O. Box 217, 7500 AE Enschede, The Netherlands
}

\date{\today}%

\begin{abstract}
In this Supplementary Online Material, we describe the technical details of the results provided in the main paper. First, we discuss how to compute capillary forces in the vicinity of the contact line within the Density Functional Theory, in case the solid and the liquid phases can be considered as perfect wedges. In this framework, we determine the equilibrium of a liquid when the solid is not a flat substrate, but a wedge. We show how Young's law and Neumann's law are recovered in particular limits. Second, we present a macroscopic linear model for the elastic response of the solid to capillary forces. This is the basis for the analytical and numerical results presented in the main paper.
\end{abstract}

\maketitle

\section{Calculation of capillary forces in wedges}

The DFT framework in which we perform our analysis separates the molecular interactions into a long-range attractive potential and a short-range hard core repulsion \cite{SnoeijerPOF2008,DasPOF2011}. The short-range repulsion is described by an internal repulsive pressure $p_{r}$ that ensures incompressibility of both the liquid and the solid --~this is a reasonable approximation for elastomers. This pressure acts as a purely normal contact force, and is continuous across the interfaces. Using Eq.~2 of the main text to define $\Phi_{SL}$ and $\Phi_{LL}$, the mechanical equilibrium allows one to express $p_r$ in the liquid as (see \cite{DasPOF2011} for the derivation):  
\begin{align}
p_{r}+\Phi_{SL}+\Phi_{LL}={}&cst\;.\label{eq:pr}
\end{align}

The force $\vec F_{LS}$ exerted by a wedge of liquid on a subsystem of the solid of volume $\mathcal V_{S}$, can be expressed as the resultant of the volumetric attraction and the repulsion at the liquid-solid interface $\mathcal S_{LS}$. Within the same framework, one can compute the force $\vec F_{SL}$ exerted by a wedge of solid on a subsystem of the liquid of volume $\mathcal V_{L}$. One writes:
\begin{align}
\vec F_{LS}={}&-\int_{\mathcal V_{S}} \vec\nabla\Phi_{LS}(r)\,d^3r-\int_{\mathcal S_{LS}}p_{r}\vec n \,d^2r\;,\label{eq:fLS}\\
\vec F_{SL}={}&-\int_{\mathcal V_{L}} \vec\nabla\Phi_{SL}(r)\,d^3r-\int_{\mathcal S_{LS}}p_{r}\vec n \,d^2r\;.\label{eq:fSL}
\end{align}
When the liquid and the solid domains are perfect wedges, these integrals can be evaluated explicitly, and expressed in terms of the surface tensions $\gamma,\gamma_{SL},\gamma_{SV}$. This will be derived below and lead to the equations of the main paper, in particular Eqs.~3, 5 and 6.
\begin{figure}[htb!]
\begin{center}
\includegraphics{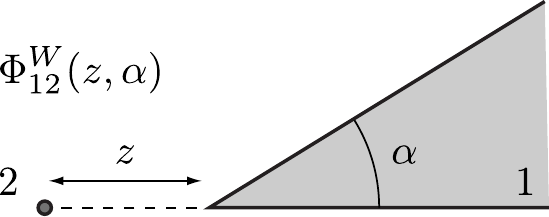}
\caption{Definition of the attractive potential $\Phi^{W}_{12}(z,\alpha)$, created by a wedge (of angle $\alpha$) of phase 1 on a point of phase 2. The expression is given in (\ref{eq:phiW}).\label{fig:phiw}}
\end{center}
\end{figure}

First, we determine the equilibrium shape of a liquid wedge on a given solid wedge, and the force exerted by the liquid on the solid in the vicinity of the contact line. This calculation can be reduced entirely in terms of the interactions between wedges, as in Fig.~\ref{fig:phiw}. The attractive potential $\Phi^{W}_{12}(z,\alpha)$ created by a wedge of a phase $1$, and characterized by an angle $0<\alpha<\pi/2$, on a point of phase 2, at a distance $z$ from the corner, and aligned with one of the two edges (see Fig.~\ref{fig:phiw}) can be analytically expressed as:
\begin{align}
\Phi^{W}_{12}(z,\alpha)=\int_{z}^\infty r\left[\alpha-\arcsin\left(\frac zr\sin\alpha\right)\right]\tilde\varphi_{12}(r)dr\;.\label{eq:phiW}
\end{align}
where $\tilde\varphi_{12}(r)=\rho_{1}\rho_{2}\int_{-\infty}^\infty\varphi_{12}(\sqrt{r^2+y^2})dy$ is the potential created by an infinite line at a distance $r$ from it.
Note that by construction, this function must be even with respect to the angle.
The more general case when a point is not aligned with an edge of the wedge can be easily deduced from the above case by adding (or subtracting) two different wedges.
Furthermore, as the surface tensions are calculated from the interactions between a column and half a space, we can obtain certain important relations governing $\tilde\varphi_{12}(r)$~\cite{Israelachvili}:
\begin{align}
\int_{0}^\infty r^2\tilde\varphi_{LL}(r)dr={}&-\gamma\;,\\
\int_{0}^\infty r^2\tilde\varphi_{LS}(r)dr=\int_{0}^\infty r^2\tilde\varphi_{SL}(r)dr={}&-\frac{\gamma+\gamma_{\rm SV}-\gamma_{\rm SL}}2=-\gamma\frac{1+\cos\theta_{Y}}2\;.
\end{align}
Here $\theta_Y$ is the Young' angle. These two integrals can now be used in (\ref{eq:phiW}) to obtain:
\begin{align}
\int_{0}^\infty \Phi^{W}_{LL}(z,\alpha)dz={}&- \gamma\frac{1-\cos\alpha}{\sin\alpha}\;,\label{eq:intpiWLL}\\
\int_{0}^\infty \Phi^{W}_{LS}(z,\alpha)dz=\int_{0}^\infty \Phi^{W}_{SL}(z,\alpha)dz={}&-\gamma\frac{1+\cos\theta_{Y}}2\frac{1-\cos\alpha}{\sin\alpha}\;.\label{eq:intpiWLS}
\end{align}
The attractive potential $\Phi_{12}^{H}(z)$ created by a half-space of a phase 1, on a point at a distance $z$ from its flat boundary can be expressed independently or by using two wedges of 90$^\circ$ :
\begin{align}
\Phi_{12}^H(z)=2\int_{z}^\infty r \arccos\frac zr\tilde\varphi_{12}(r)\,dr=2\,\Phi^W_{12}\left(z,\frac\pi2\right)\;.
\end{align}
This quantity (noted $\Pi_{12}(z)$ in \cite{DasPOF2011}) can be interpreted as a disjoining pressure. Furthermore, we have some useful symmetry properties that we can easily interpret from geometrical arguments (here we dropped the subscripts to clarify the notations):
\begin{align}
\Phi^H(z<0)={}&2\Phi^H(0)-\Phi^H(-z)\,,\label{eq:sym1}\\
\Phi^W(z,\alpha>\pi/2)={}&\Phi^H(z\sin\alpha)-\Phi^W(z,\pi-\alpha)\;,\label{eq:sym2}\\
\Phi^W(z<0,\alpha)={}&\Phi^H(0)-\Phi^W(-z,\pi-\alpha)\;.\label{eq:sym3}
\end{align}

\subsection{Force exerted by the solid on the liquid near the contact line}

We denote by $\mathcal S_{LV}$, $\mathcal S_{SV}$ and $\mathcal S_{SL}$ the liquid-vapor, solid-vapor and solid-liquid interfaces, respectively. The subsystem of liquid  subject to forces is bounded on the liquid side by a surface perpendicular to the liquid-solid interface, denoted $\mathcal S_{LL}$ (see Fig.~\ref{fig:wedgeL}).
The angle of the liquid wedge is $0<\theta_{L}<\pi/2$ and the angle of the solid wedge is $0<\theta_{S}<\pi$.
The force exerted by the solid wedge on the liquid subsystem in the vicinity of the corner is given by~(\ref{eq:fSL}). Using Ostrogradsky's theorem, the volume integral in (\ref{eq:fSL}) reduces to a surface integral, in terms of the surfaces bounding the volume of the liquid subsystem:
\begin{align}
\vec{F}_{SL}={}&-\int_{\mathcal S_{LV}}\Phi^{(1)}_{SL}\,\left(-\sin\theta_L\vec u_{x}+\cos\theta_L\vec u_{z}\right) dS
-\int_{\mathcal S_{LL}}\Phi^{(2)}_{SL}\,\vec u_{x}\,dS
+\int_{\mathcal S_{SL}} \left(\Phi^{(3)}_{SL}+p_{r}^{(3)}\right)\vec u_{z}\, dS\;,\label{eq:ForceSL}
\end{align}
where $\vec u_{x}$ and $\vec u_{z}$ are the tangential and the normal unit vectors to the solid-liquid interface, both directed toward the interior of the liquid. We thus have to calculate the three different attractive potentials $\Phi^{(i)}_{SL}$ created by the solid wedge on different locations of the liquid wedge, and the repulsive pressure $p^{(3)}_{r}$ in the liquid at the liquid-solid interface. Let us first concentrate on the liquid at the liquid-solid interface. The liquid is at equilibrium, so we can use Eq.~(\ref{eq:pr}) to obtain (in the entire liquid) $p_{r}+\Phi_{SL}+\Phi_{LL}=\Phi^H_{LL}(0)$, which can be simplified to obtain:
\begin{align}
p_{r}^{(3)}+\Phi^{(3)}_{SL}=\Phi^W_{LL}(x,\pi-\theta_{L})\;.\label{eq:phi3}
\end{align}
Then, we can express the solid-liquid attractive potentials $\Phi^{(2)}_{SL}$ and $\Phi^{(1)}_{SL}$ respectively far from the contact line at a distance $z$ from the solid-liquid interface, and along the liquid-vapor interface at a distance $s$ from the contact line:
\begin{align}
\Phi^{(2)}_{SL}={}&\Phi^H_{SL}(z)\label{eq:phi2}\;,\\
\Phi^{(1)}_{SL}={}&\Phi^W_{SL}(s,\pi-\theta_{L})+\Phi^W_{SL}(s,\theta_{L}+\theta_{S}-\pi)\;.\label{eq:phi1}
\end{align}
Using the different expressions of the potentials above (\ref{eq:phi3}-\ref{eq:phi1}) and the integrals (\ref{eq:intpiWLL},\ref{eq:intpiWLS}), we can finally explicitly express the force (\ref{eq:ForceSL}) per unit $L$ of contact line along the two directions normal and tangential to the solid-liquid interface:
\begin{align}
f^{t}_{SL}=\frac{\vec F_{SL}\cdot\vec u_{x}}{L}
={}&\gamma\left(1+\cos\theta_{Y}\right)\frac{\sin\theta_{L}}2\left[\frac{1-\cos\theta_{L}}{\sin\theta_{L}}+\frac{1+\cos(\theta_{L}+\theta_{S})}{\sin(\theta_{L}+\theta_{S})}\right]\;,\label{eq:fSLt}\\
f^{n}_{SL}=\frac{\vec F_{SL}\cdot\vec u_{z}}{L}
={}&\gamma\left[\left(1+\cos\theta_{Y}\right)\frac{\cos\theta_{L}}2\left(\frac{1+\cos\theta_{L}}{\sin\theta_{L}}-\frac{1+\cos(\theta_{L}+\theta_{S})}{\sin(\theta_{L}+\theta_{S})}\right)-\frac{1+\cos\theta_{L}}{\sin\theta_{L}}\right]\;.\label{eq:fSLn}
\end{align}

\subsection{Liquid equilibrium}
\begin{figure}[t!]
\begin{center}
\includegraphics{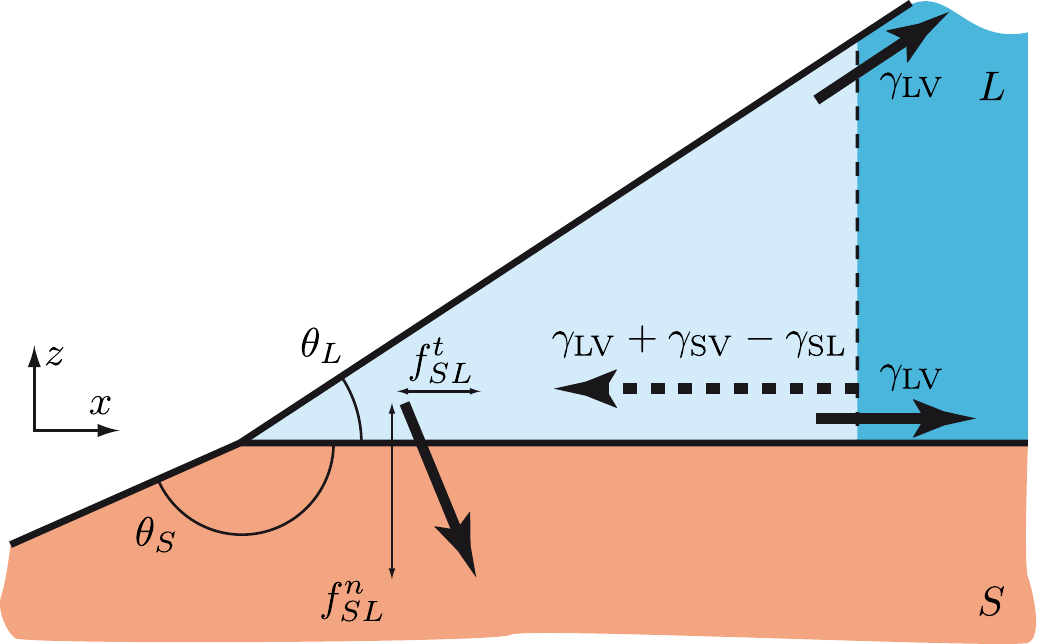}
\caption{Distribution of capillary  forces induced in a subsystem of the liquid wedge near the contact line.\label{fig:wedgeL}}
\end{center}
\end{figure}
As explained in the main text, the equilibrium condition for the liquid is obtained by balancing the solid-on-liquid force with the forces due to liquid-liquid interactions. The system on which the forces act is taken as the liquid corner depicted in Fig.~\ref{fig:wedgeL}). The liquid-liquid interactions can be separated in two contributions. First, there are attractive liquid on liquid forces, which act at the liquid-liquid and the solid-liquid interfaces, far from the contact line. Each of these two forces are equal to $\gamma$ per unit contact line, and are pulling on these two corners along the liquid-vapor and liquid-solid interfaces. The second is due to repulsive liquid-liquid interactions, acting near the boundary normal to the solid-liquid interface far from the contact line. It originates from the presence of the attractive solid. The resulting tangential force (per length unit of contact line) is $\gamma+\gamma_{\rm SV}-\gamma_{\rm SL}=\gamma(1+\cos\theta_{Y})$ and is directed toward the interior of the subsystem \cite{MarchandAJP11}. The balance of these forces, tangential and normal to the interface, read:
\begin{align}
\gamma(1+\cos\theta_{L})-\gamma\left(1+\cos\theta_{Y}\right)+f^t_{SL}={}&0\;,\label{eq:eqLt}\\
\gamma\sin\theta_{L}+f^n_{SL}={}&0\;.\label{eq:eqLn}
\end{align}
These two equations come from the projections of the gradient of a scalar: it implies that they represent a single independent equation that reduces to [using for instance (\ref{eq:fSLt}) and (\ref{eq:eqLt})]:
\begin{align}
1+\cos\theta_{Y}=\left(1+\cos\theta_{L}\right)\left(1+\frac{\tan(\theta_{L}/2)}{\tan(\theta_{S}/2)}\right)\;.
\end{align}
The unique physical solution of this equation is given by:
\begin{align}
\cos\theta_L=\frac12\left[ \cos\theta_Y[1-\cos\theta_S]-\sin\theta_S \sqrt{ \frac2{1-\cos\theta_S} -\cos^2\theta_Y } \right]\;,
\label{eq:thetaL}
\end{align}
which is Eq.~(3) in the main paper. 

\subsection{Young's law and Neumann's law}
One easily shows that in the case of a flat substrate, $\theta_{S}=\pi$, the tangential force $f^t_{SL}$ vanishes due to symmetry reasons: the solid spans an infinite half-space in the tangential direction. We then recover $\theta_{L}=\theta_{Y}$, [either from (\ref{eq:eqLt}) or  (\ref{eq:thetaL})], which is Young's law. In addition we note that $f^n_{SL}=\gamma\sin\theta_{Y}$, ensuring that we have a force balance in the liquid subsystem in both directions.

By assuming the substrate is a perfect liquid, we will now show that the above formulation is equivalent to Neumann's law~\cite{Neumann}. Denoting this liquid substrate as phase $2$, and the liquid wedge on top of it as phase $1$, Neumann's law is given by the two following equations:
\begin{align}
\gamma_{1}\sin\theta_{1}-\gamma_{2}\sin\theta_{2}={}&0\;,\label{eq:neumann1}\\
\gamma_{1}\cos\theta_{1}+\gamma_{2}\cos\theta_{2}+\gamma_{12}={}&0\;,\label{eq:neumann2}
\end{align}
where $\gamma_{1}$ and $\gamma_{2}$ are the surface tensions of liquids 1 and 2, $\gamma_{12}$ is the interfacial tension, and $(\theta_{1}, \theta_{2})$ are the positive angles made by the liquid wedges 1 and 2. Neumann's law is in fact ruled by two independent parameters, which we denote according to Young's law by:
\begin{align}
\cos\theta_{Y1}={}&\frac{\gamma_{2}-\gamma_{12}}{\gamma_{1}}\;,\label{eq:young1}\\
\cos\theta_{Y2}={}&\frac{\gamma_{1}-\gamma_{12}}{\gamma_{2}}\;.\label{eq:young2}
\end{align}
With these notations, we  get $\gamma_{1}/\gamma_{2}=(1+\cos\theta_{Y2})/(1+\cos\theta_{Y1})$ and $\gamma_{12}/\gamma_{2}=(1-\cos\theta_{Y1}\cos\theta_{Y2})/(1+\cos\theta_{Y1})$, 
and Neumann's law (\ref{eq:neumann1},\ref{eq:neumann2}) can be rewritten as:
\begin{align}
\sin\theta_{1}(1+\cos\theta_{Y2})-\sin\theta_{2}(1+\cos\theta_{Y1})={}&0\;,\\
\cos\theta_{1}(1+\cos\theta_{Y2})+\cos\theta_{2}(1+\cos\theta_{Y1})={}&\cos\theta_{Y1}\cos\theta_{Y2}-1\;,
\end{align}
The unique physical solution can then be explicitly expressed as:
\begin{align}\label{eq:explicitneumann}
\cos\theta_{1}=\frac{ \big[2+(1-\cos \theta_{Y2})\cos \theta_{Y1} \big]\cos \theta_{Y1}-(1+\cos \theta_{Y2})}{2 (1-\cos \theta_{Y1}\cos \theta_{Y2})}\;.
\end{align}
or equivalently by a function of the form $\theta_{1}=N(\theta_{Y1},\theta_{Y2})$. The second angle is given by symmetry, i.e., $\theta_{2}=N(\theta_{Y2},\theta_{Y1})$.

Let us now show that the DFT formulation indeed reduces to Neumann's law when both the phases are liquid. For a given substrate wedge of angle $\theta_{2}$, we can see from (\ref{eq:thetaL}) that the liquid 1 is at equilibrium when it forms a wedge of angle $\theta_{1}=f(\theta_{Y},\theta_{2})$. If the phase 2 is a liquid as well, then we can use the same equation (\ref{eq:thetaL}) with exchanged roles of phase 1 and 2. This gives two equations for the contact angles:
\begin{align}
\theta_{1}={}&f(\theta_{Y1},\theta_{2})\;,\\
\theta_{2}={}&f(\theta_{Y2},\theta_{1})\;,
\end{align}
where $\theta_{Y2}$ is defined in (\ref{eq:young2}). Therefore, $\theta_{1}$ obeys the following implicit equation:
\begin{align}
\theta_{1}={}&f\big(\theta_{Y1},f(\theta_{Y2},\theta_{1})\big)\;.
\end{align}
We can easily verify that this equation has a unique solution in the interval $[0,\pi]$, which is given explicitly by (\ref{eq:explicitneumann}). By analogy, we also recover the second angle $\theta_{2}$ with the symmetry $\theta_{Y1}|\theta_{Y2}$.

\subsection{Force exerted by the liquid on the solid near the contact line}

Contrarily to the liquid, the capillary forces on the solid need not be at equilibrium for the solid: a resultant force can be balance by the elasticity. By analogy with the determination of the solid on liquid force, we compute the capillary force (\ref{eq:fLS}) exerted by the entire wedge of liquid to a subsystem of solid bounded by a surface perpendicular to the liquid-solid interface, noted $\mathcal S_{SS}$:
\begin{align}
\vec{F}_{LS}={}&-\int_{\mathcal S_{SV}}\Phi^{(1)}_{LS}\,\left(-\sin\theta_S\vec u_{x}-\cos\theta_S\vec u_{z}\right) dS
-\int_{\mathcal S_{SS}}\Phi^{(2)}_{LS}\,\vec u_{x}\,dS
-\int_{\mathcal S_{SL}} \left(\Phi^{(3)}_{LS}+p_{r}^{(3)}\right)\vec u_{z}\, dS\;.\label{eq:ForceLS}
\end{align}
The attractive potentials $\Phi^{(1)}_{LS}=\Phi^W_{LS}(s,\pi-\theta_{S})+\Phi^W_{LS}(s,\theta_{L}+\theta_{S}-\pi)$ and $\Phi^{(2)}_{LS}=\Phi^H_{LS}(z)$ are analogous to what we obtained in (\ref{eq:phi1}) and (\ref{eq:phi2}). The only fundamental difference is that for this case the pressure is determine by the liquid (and not the solid) equilibrium (\ref{eq:phi3}), so that:
\begin{align}
p_{r}^{(3)}+\Phi^{(3)}_{LS}=\Phi^W_{LL}(x,\pi-\theta_{L})+\Phi^W_{SL}(x,\pi-\theta_{S})-\Phi^W_{LS}(x,\pi-\theta_{L})\;.\label{eq:phiLS3}
\end{align}
Therefore, after some simplifications, we can determine [from (\ref{eq:ForceLS})] the normal and the tangential components of the liquid on solid force (per unit length of the contact line):
\begin{align}
f^{t}_{LS}=\frac{\vec F_{LS}\cdot\vec u_{x}}{L}\label{eq:fLSt}
={}&\gamma\frac{1+\cos\theta_{Y}}{2}\left[1-\cos\theta_{S}+\sin\theta_{S}\frac{1+\cos(\theta_{L}+\theta_{S})}{\sin(\theta_{L}+\theta_{S})}\right]\;,\\
f^{n}_{LS}=\frac{\vec F_{LS}\cdot\vec u_{z}}{L}
={}&\gamma\left[\frac{1+\cos \theta_{Y}}{2} \left(\sin \theta_{S}+\cos\theta_{S} \frac{1+\cos (\theta_{L}+\theta_{S})}{ \sin(\theta_{L}+ \theta_{S})}\right)+\frac{1-\cos \theta_{Y}}{2}\frac{1+\cos \theta_{L}}{ \sin \theta_{L}} \right]\;.\label{eq:fLSn}
\end{align}
These expressions can be rewritten in a more compact form as Eqs.~(5) and (6) of the main paper. When $\theta_{S}=\pi$, then $\theta_{L}=\theta_{Y}$ [using (\ref{eq:thetaL})] and $f^{n}_{LS}=\gamma\sin\theta_{Y}$ [using (\ref{eq:fLSn})], as we would expect from the vertical action-reaction interaction of two full wedges, and $f^{t}_{LS}=\gamma(1+\cos\theta_{Y})$ [using (\ref{eq:fLSt})], as was shown in \cite{DasPOF2011,MarPRL2012}.

\begin{figure}[t!]
\begin{center}
\includegraphics{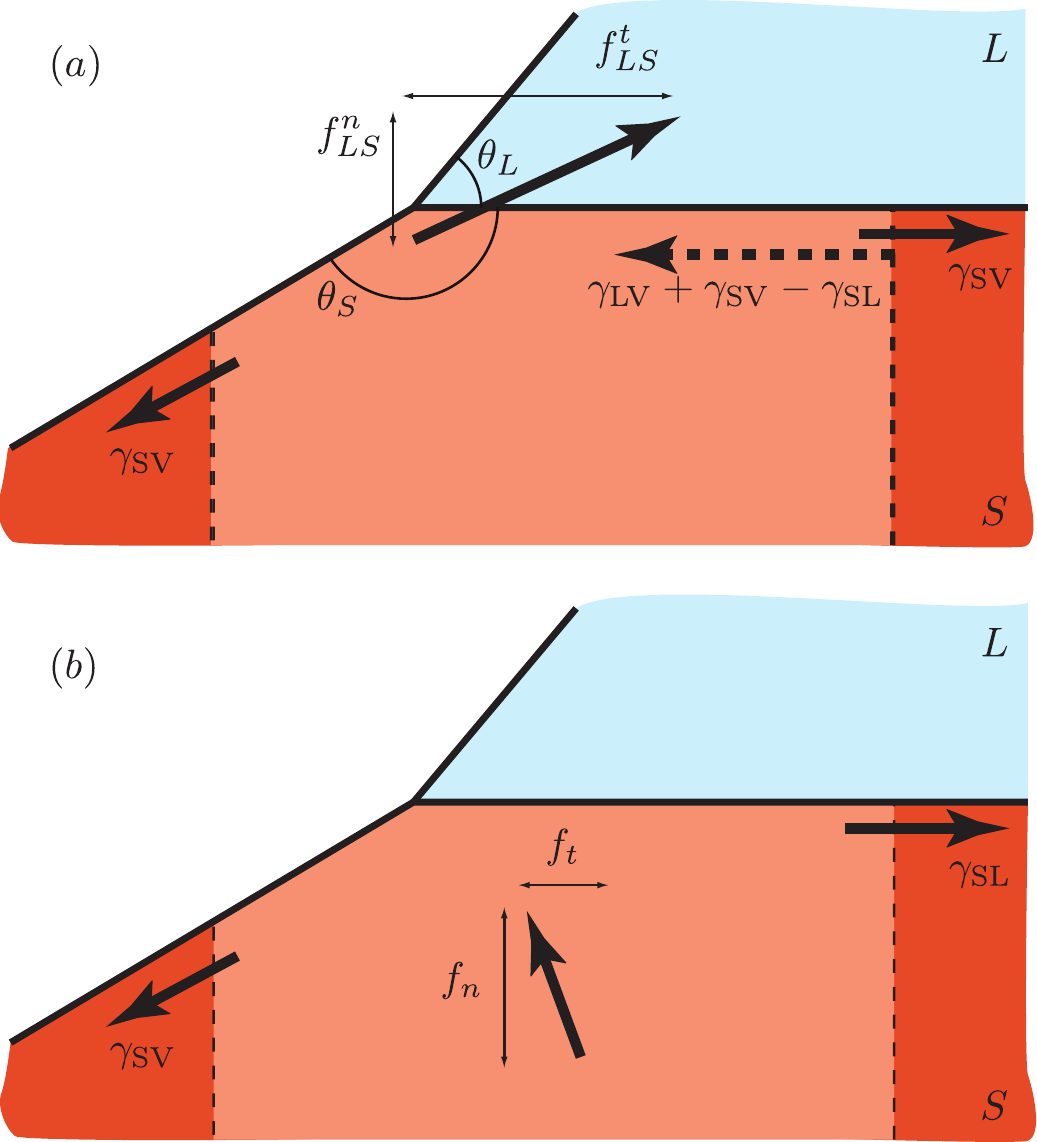}
\caption{(a) Distribution of capillary induced forces inside the solid. (b) Equivalent representation allowing a distinction between the forces at the contact line, and the forces that could be compensated by a Laplace pressure when the interfaces are curved.\label{fig:wedgeS}}
\end{center}
\end{figure}

\section{Macroscopic elasto-capillary model}
The description of the solid is more subtle than that of the liquid. The bulk of the elastic substrate is described as a homogeneous incompressible elastic medium. However, one needs to define the effective boundary conditions to be applied to the bulk elasticity equations to compute the deformations. As discussed in the main text, one can consider two models that will be referred to as \emph{normal force transmission} and \emph{vectorial force transmission} respectively. Below we present the expressions for the surface stress $\sigma_s$ used for our numerical results.

\subsection{Curved interfaces interactions}
For later reference, we first determine the attractive potential at an interface between two phases 1 and 2 which presents a curvature $\kappa$. For $\kappa a\ll1$, this potential (Eq.~2 in the main text) reduces to:
\begin{align}
\Phi_{12}=\Phi_{12}^H(0)-\kappa\int_{0}^\infty r^2\tilde\varphi_{12}(r)dr\;,\label{eq:laplace}
\end{align}
where, $\kappa$ is defined as positive when phase 1 is smaller than phase 2. In (\ref{eq:laplace}), the integral term is always negative, and depending on whether one considers liquid-liquid, solid-solid or liquid-solid interaction, becomes equal to $-\gamma$, $-\gamma_{\rm SV}$ or $-\gamma(1+\cos\theta_{Y})/2$ [see (\ref{eq:intpiWLL},\ref{eq:intpiWLS})]. The Laplace pressure originates from this potential.

\subsection{Normal force transmission model}
An elastomer can be described as an incompressible soft solid within which the elastic stress can be written as:
\begin{equation}
\sigma_{ij}=-p_r \delta_{ij}+\frac E3\;\left(\partial_i u_j+\partial_j u_i\right)\;,
\end{equation}
where $u_i$ is the displacement vector, and $E$ the Young's modulus. This solid is assumed to be submitted to the long-range attractive potential. Then it can be rigorously shown (see the proof below) that the the external solution must have an effective boundary condition: 
\begin{equation}\label{eq:hap}
\vec \sigma_s=\left(\Phi_{LS}+\Phi_{SS}-\Phi_{SL}-\Phi_{LL}+\Phi^H_{LL}(0)-\Phi^H_{SS}(0)\right)\vec n\;.
\end{equation}
Note that this is the opposite of the surface force. When the liquid phase is expressed by a vapor phase, this expression reduces to:
\begin{equation}
\vec \sigma_s=\left(\Phi_{SS}-\Phi_{SS}^H(0)\right)\vec n\;.\label{eq:sigma_s_SS}
\end{equation}
The numerical solution presented in the main paper are based on numerical evaluations of the surface stress (\ref{eq:hap},\ref{eq:sigma_s_SS}).

The surface stress $\sigma_s$ acts as a generalized capillary pressure. At distances from the contact line much greater than the molecular cut-off $a$, the curvature is small so that [using (\ref{eq:laplace},\ref{eq:sigma_s_SS})] the capillary pressure on the dry part reduces to the standard Laplace pressure:
\begin{equation}
\vec \sigma_s =\gamma_{SV}\;\kappa\;\vec n\;. \label{eq:laplaceSVll}
\end{equation}
On the other hand, the Laplace pressure on the wet part is 
\begin{equation}
\vec \sigma_s=\gamma_{SL}\;\kappa\;\vec n\;.\label{eq:laplaceSLll}
\end{equation}

\textbf{Proof~--~} The incompressibility condition reads $\partial_l u_l=0$ or in vector notations:
$$\vec \nabla\cdot\vec u=0.$$
We first assume that the constitutive equation holds in the interfacial zone where the elastomer is subjected to two long range volumetric interactions which derive from the potentials $\Phi_{LS}$ and $\Phi_{SS}$ that are associated to liquid-solid and solid-solid interactions respectively. The equilibrium condition reads:
\begin{equation}
-\vec \nabla(p_r+\Phi_{LS}+\Phi_{SS})+E\vec \nabla^2 \vec u=0.
\end{equation}
$\Phi_{LS}$ and $\Phi_{SS}$ result from an integration over the volume of the liquid and the solid. They vary over a scale of few molecular sizes along the normal to the solid interface and attain their bulk values at distances away from the interface. Therefore, $\Phi_{LS}$ and $\Phi_{SS}$ are only important in a thin boundary layer across the interface. We pose $\tilde p=p_r+\Phi_{LS}+\Phi_{SS}$ and $\tilde \sigma_{ij}=\sigma_{ij}-(\Phi_{LS}+\Phi_{SS}) \delta_{ij}$. Then the equation reduces to:
\begin{equation}
-\vec \nabla \tilde p+E\vec \nabla^2 \vec u=0
\end{equation}
in the whole domain. This is the standard elasticity problem. However the effective boundary conditions are modified. More precisely, there is no excess quantity in the strain tensor, but the effective stress applied to the bulk is modified. As the real stress $\sigma_{ij}$ is continuous, the pseudo-stress $\tilde \sigma_{ij}$ is not.

\subsection{Vectorial force transmission model}
We can alternatively propose a model of elastomeric surface that allows for a fully vectorial transmission of the liquid-on-solid force. The polymeric chains are attracted in the direction of the liquid and thus transfer the total force exerted by the liquid on the solid. There is no force gradient building up in the surface layer. Integrating the equilibrium equation from the surface to the bulk, we see that the effective stress perfectly balances the excess forces due to the long range interactions:
\begin{eqnarray}
\vec \sigma_s&=&\vec t\cdot \vec \nabla \left[\int_0^\infty \Phi_{LS} dh\;\vec t\right]  \\
&+&\left(\Phi_{LS}+\Phi_{SS}-\Phi_{SL}-\Phi_{LL}+\Phi^H_{LL}(0)-\Phi^H_{SS}(0)\right)\vec n \nonumber
\end{eqnarray}
This is the expression for $\sigma_s$ used in the numerical solution of the vectorial force transmission model. 

Again, looking at a distance from the contact line large compared to $a$ and using
$$\vec t\cdot \vec \nabla \vec t=-\kappa \vec n,$$
we get in the dry region a surface stress:
\begin{equation}
\vec \sigma_s=\left(\gamma_{SV}+\gamma\right)\;\kappa\;\vec n
\end{equation}
Beside, the Laplace pressure in the dry part is still 
\begin{equation}
\vec \sigma_s =\gamma_{SV}\;\kappa\;\vec n. 
\end{equation}

\bibliography{DFTPRL_suppl}